\documentstyle[11pt,newpasp,twoside,epsf]{article}
\def\etal{{et~al.\ }}

\def\eg{e.g.,\ }

\def\m5007{m$_{5007}$}
\def\kms{km s$^{-1}$}

\def\edcomment#1{\iffalse\marginpar{\raggedright\sl#1\/}\else\relax\fi}
\reversemarginpar

\begin{document}
\title{Intracluster Planetary Nebulae}

\author{John J. Feldmeier}
\affil{Case Western Reserve University, 10900 Euclid Ave. Cleveland, 
OH 44106, U.S.A.}

\begin{abstract}
We review the progress of research on intracluster planetary 
nebulae (IPN).  In the past five years, hundreds of IPN candidates 
have been detected in the Virgo and Fornax galaxy clusters and  
searches are also underway in poorer galaxy groups.  From the 
observations to date, and applying the known properties of 
extragalactic planetary nebulae, the intracluster light in Virgo
and Fornax: 1) is significant, at least 20\% of the total cluster 
stellar luminosity, 2) is elongated in Virgo along our line of sight, 
and 3) may derive from lower-luminosity galaxies, consistent with
some models of intracluster star production.  A fraction of
IPN candidates are not true IPN, but emission-line sources
of very large observed equivalent width ($\geq 200$ \AA).  
The most likely source for these contaminating 
objects are Lyman-$\alpha$ galaxies at z $\approx$ 3.1.  
Follow-up spectroscopy of the IPN candidates will be crucial 
to discriminate against high redshift galaxies and to derive the
velocity field of the intracluster stellar population.

\end{abstract}

\section{Introduction}
Intracluster stars, stars between the galaxies in a galaxy cluster, 
are an excellent probe of galaxy and galaxy cluster evolution.
They preserve a record of a galaxy cluster's dynamical history, and
are a sensitive measure of the poorly understood processes 
of galactic mergers, cluster accretion, and tidal-stripping 
that occur in galaxy clusters (cf.~Dressler 1984).  

The original concept of intracluster stars was first proposed 
over fifty years ago (Zwicky 1951), but actually 
observing intracluster starlight has been extremely difficult.  
Typically, the surface brightness of intracluster starlight 
is less than 1\% that of the brightness of the sky background, 
and measurements of this luminosity must contend with the 
problems presented by scattered light from bright objects and 
the contribution of discrete sources below the detection limit.
These difficulties have made research into intracluster 
starlight slow and uncertain, though solid detections have been 
made (for reviews, see V{\'i}lchez-G{\'o}mez 1999; Feldmeier 2000).

An alternative way to study intracluster light is to search for it
in nearby galaxy clusters.  Here, it is possible to search for 
luminous individual intracluster stars, and gain more detailed 
information on the distribution, metallicity and velocities 
of intracluster stars than is possible from surface brightness
measurements.  Planetary nebulae (PN) are one such 
tracer of the intracluster starlight.   

The history of intracluster planetary nebulae (IPN) research 
begins a decade ago with the first PN survey of the 
Virgo cluster (Jacoby, Ciardullo, \& Ford 1990; hereafter JCF).  In this
survey of elliptical galaxies, JCF found 11 PN 
that were much brighter than the expected [O~III] $\lambda$ 5007 
planetary nebulae luminosity function (PNLF) cut-off magnitude.  JCF attempted
to explain these ``overluminous'' PN with a number of hypotheses, 
but none was entirely satisfactory.

The next step involved spectroscopic follow-up of objects from the JCF
survey.  During a radial velocity survey of PN in the Virgo 
elliptical galaxy M~86, Arnaboldi \etal (1996) found that 16 
of the 19 detected PN velocities were consistent with the galaxy's mean
velocity (v$_{radial}$ = -227 \kms).  The other three planetaries
had mean velocities of $\sim 1600$ \kms, more consistent with
the Virgo cluster's mean velocity.  Arnaboldi \etal (1996) 
argued convincingly that these objects were intracluster planetary 
nebulae, and it is here that the term first enters the literature.  
Almost simultaneously, the first search for IPN candidates in
the Fornax cluster was published (Theuns \& Warren 1997), and more 
detections of IPN candidates in Virgo quickly followed 
(M\'endez \etal 1997; Ciardullo \etal 1998; Feldmeier, 
Ciardullo, \& Jacoby 1998).  Additionally, Ferguson, Tanvir and
von Hippel (1998) detected intracluster red giant stars (IRGs) in the Virgo
cluster using {\sl HST}.  This provided independent 
confirmation of the IPN results, and allowed for a potential 
comparison between the two detection methods.

With the help of large CCD mosaics, samples of hundreds of 
IPN have now been gathered in the Virgo and Fornax clusters (Fig 1., 
Feldmeier 2000; Arnaboldi \etal 2002).  IPN candidates are easily 
identified as stellar sources that appear in a deep [O~III] 
$\lambda$ 5007 image, but completely disappear in an image through a
filter that does not contain the [O~III] line.  Originally, 
searching for such objects was done manually, 
but now is done through automated methods (Feldmeier 2000; 
Arnaboldi \etal 2002)
\begin{figure}
{\vskip100pt}
\caption{Images of portions of the Virgo (top) and Fornax
(bottom) galaxy clusters, with the locations of IPN and other 
intracluster star detections marked.  
There are 319 IPN candidates found in Virgo Fields 1-8, and 
138 candidates in the Fornax fields.  Virgo's subclumps A \& B
are centered at the the top and bottom of the Virgo image.}
\end{figure}

\section{Why study IPN?}  
IPN are a useful tracer of intracluster stars for a number of
reasons.  First, since PN are a normal end-phase of stellar evolution, 
the distribution of PN in a stellar population should closely 
follow the spatial distribution of normal stars.  Second, because 
a large portion ($\sim 15\%$) of the total light of a bright PN is
emitted in a single emission line at 5007 \AA, radial
velocities of the IPN can be obtained, and dynamical information
gained.  The utility of planetary nebulae as dynamical probes 
will be covered by others (see Arnaboldi, Freeman, Mathieu, 
this conference), but it's important to stress that the only way to 
obtain dynamical information about intracluster starlight is through IPN 
observations.  Finally, since PN follow a well-determined 
luminosity function, it is possible to obtain some 
information on the three-dimensional structure of 
intracluster stars.

For the remainder of this paper, we will focus on obtaining 
the amount of intracluster luminosity using the IPN as a tracer.  
This aspect directly compares to models of intracluster starlight, and 
is the area where the greatest progress has been made thus far.

\section{Obtaining an intracluster luminosity from IPN}
In principle, determining the amount of intracluster luminosity from 
the observed numbers of IPN is straightforward.  Theories of simple 
stellar populations (\eg Renzini \& Buzzoni 1986) has shown that the 
bolometric luminosity-specific stellar evolutionary flux 
of non-star-forming stellar populations should 
be $\sim 2 \times 10^{-11}$~stars-yr$^{-1}$-$L_{\odot}^{-1}$, nearly 
independent of population age or initial mass function.  If the 
lifetime of the planetary nebula stage is $\sim 25,000$~yr, then 
every stellar system should have $\alpha \sim 50 \times 
10^{-8}$~PN-$L_{\odot}^{-1}$.  If we also assume that the 
[O~III] $\lambda$ 5007 PNLF follows the well studied form:  
$N(M) \propto e^{0.307M_{5007}} \, [1 - e^{3(M_{5007}^{*}-M)}]$
(Ciardullo \etal 1989), we can determine the total expected number of IPN 
in a region, and therefore, the luminosity, from observations.  
Due to observational constraints, we calibrate our IPN 
observations using the quantity $\alpha_{2.5}$, the number of 
PN up to 2.5 magnitudes below the luminosity function cutoff, per 
stellar bolometric luminosity.  The $\alpha_{2.5}$
parameter is approximately one-tenth of $\alpha$, or $\alpha_{2.5} 
\sim 50 \times 10^{-9}$~PN-$L_{\odot}^{-1}$

In practice, the process is more complicated.  There are several  
effects that must be accounted for before the best estimate of 
intracluster luminosity can be obtained.  Fortunately, each of these 
effects also gives us an opportunity to learn more about the properties 
of the underlying intracluster stellar population.  We now discuss 
each of these effects.

\section{Line-of-sight effects}
In a distant galaxy, theory and observations show that PN follow a
well-defined [O~III] luminosity function (Ciardullo, this conference).  
However, in the galaxy cluster environment IPN are not found at 
the same distance: they are spread out substantially along our 
line-of-sight.  This distorts the observed luminosity function and 
introduces a significant uncertainty in the derived amount of 
intracluster luminosity.  From modeling (Feldmeier, Ciardullo, \& 
Jacoby 1998), we have found that the amount of intracluster 
luminosity found for a particular region can
vary by a factor of three between a model where all IPN lie at
the same distance, and a model where they are distributed uniformly
in a spherical configuration.  Therefore, to determine the true amount of
intracluster starlight, we must also learn the line-of-sight density
of the IPN.  

Fortunately, the line-of-sight density of IPN is of considerable
scientific interest.  Figure 2 shows the distance to
each IPN observed field in Virgo compared to that found from the
galaxies, assuming that the brightest IPN candidate in each field is 
near the PNLF cutoff magnitude, M$^{*}_{5007}$ = -4.48.  The most 
obvious feature of the IPN upper limit distances is that 
there is a noticeable systematic offset in distances between IPN fields in 
subclumps A \& B in Virgo.  This is most likely due to the 
structure of the Virgo cluster.  Observations by 
Yasuda, Fukugita, \& Okamura (1997) and others find that the 
galaxies of subclump~B have a mean distance modulus $\sim 0.4$~mag 
more distant than those of subclump A{}, in agreement with the upper
limit distances shown here. 

The next most obvious feature is the relatively small distance 
to all of the fields in subclump~A.  The median distance of 
these fields is approximately 12 Mpc, much smaller than the 
mean distance to Virgo found through PNLF observations of 
ellipticals and {\sl HST} observations of Cepheids.  The reason 
for this discrepancy is that all Virgo IPN surveys have a 
selection effect: objects on the near side of the cluster are 
much more likely to be found than those on the far side.  
Therefore, the 12 Mpc distance is representative of the front edge of 
the cluster, and not of the cluster core.  This result 
naturally explains the ``overluminous'' PN found by JCF.  
These PN were not overluminous at all, but instead were 
significantly closer than the other PN.  When we compare the 
cluster depth found to the angular size of Virgo,
we find that Virgo is elongated along our line of sight, perhaps
by a factor of two.  This result has been found before 
(\eg Yasuda, Fukugita, \& Okamura 1997), but to independently
confirm it through the IPN data is encouraging.
\begin{figure}
\plotfiddle{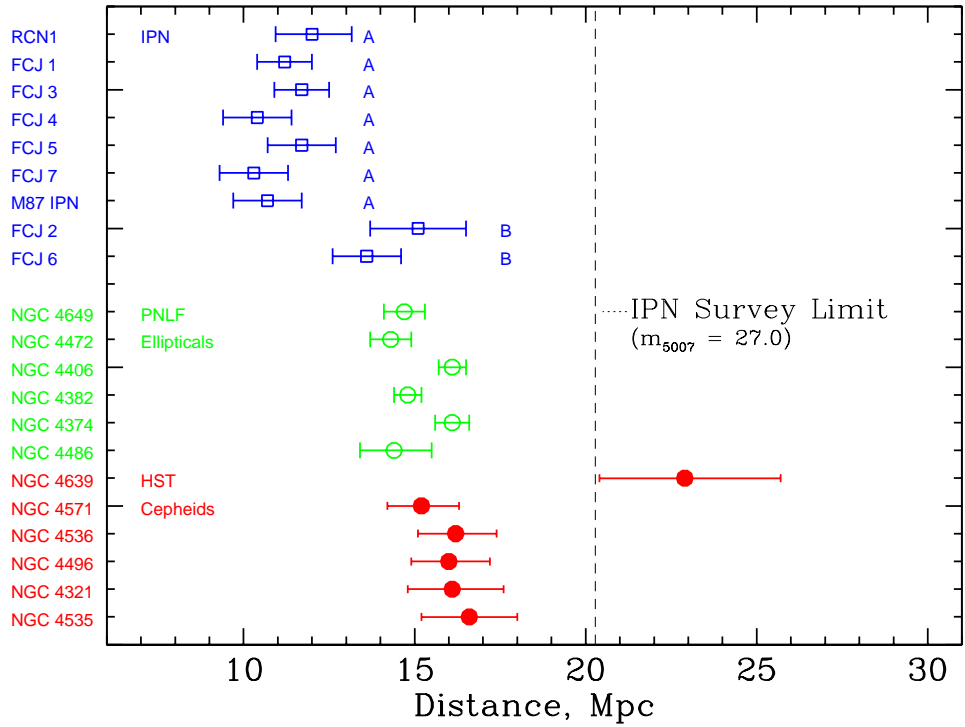}{50pt}{-90}{50}{50}{-190}{100}
{\vskip200pt}
\caption{A comparison of distances to the Virgo cluster from three samples.
At bottom are distances to spiral galaxies using {\sl HST} 
observation of Cepheid
stars, in the middle are distances to elliptical galaxies using the PNLF
method, and at top are distances to the IPN fields.  Note that there is
a clear distance offset between IPN in subclumps A and B, and the 
distances to IPN in subclump A are at least 3 Mpc in front of 
the mean distance to the Virgo cluster.}
\end{figure}
\vskip -0.5in
\section{The true production rate of IPN}
Although theory predicts a constant rate between the number of 
planetaries and bolometric stellar luminosity, observations
tell a different story.  Ciardullo (1995) found that in a 
sample of elliptical galaxies the observed 
$\alpha_{2.5}$ rate never exceeded $\alpha_{2.5} 
= 50 \times 10^{-9}$~PN-$L_{\odot}^{-1}$
but was often less, up to an order of magnitude, 
with higher luminosity galaxies having smaller values 
of $\alpha_{2.5}$.  Why this occurs is not entirely 
clear: it is believed that as a stellar population 
ages or becomes metal-rich, not all stars undergo the PN phase 
(\eg Ciardullo, Jacoby \& Feldmeier 1994)
Since the total amount of intracluster luminosity found 
is inversely proportional to $\alpha_{2.5}$, a large 
systematic error in the luminosity can be introduced if 
this effect is not accounted for.

To correct for this effect, we have made additional IRG observations of 
Virgo's core using {\sl HST} (Durrell \etal 2002).  The numbers of 
red giant stars are another determinant of the total luminosity of a 
stellar population, and therefore we can directly 
measure $\alpha_{2.5}$ from the Virgo
cluster environment.  By comparing the numbers of IPN in a field 
surrounding the {\sl HST}~field, we find a preliminary value of 
$\alpha_{2.5} = 23^{+10}_{-12} \times 10^{-9}$~PN-$L_{\odot}^{-1}$ 
(Durrell \etal 2002).  This value has an additional systematic 
uncertainty: there is evidence for a non-random spatial distribution 
in the IPN (Feldmeier 2000), and the angular sizes of the {\sl HST} field
and the IPN field differ greatly.  However, the value for $\alpha_{2.5}$
strongly implies that the bulk of the IPN came from lower-luminosity 
galaxies, and not giant ellipticals (M$_{B} \geq -20$; Ciardullo 1995).  
This is consistent with the ``galaxy harassment'' models of 
Moore \etal (1996), who predicted that most intracluster light 
would originate from lower luminosity galaxies.  

\section{Background Contamination}

Finally, not all IPN candidates that are detected photometrically are 
bona-fide IPN.  Although spectroscopic follow-up observations 
clearly shows that many IPN candidates are genuine, there are also 
a significant number of contaminating sources (Kudritzki \etal 2000; 
Freeman \etal 2000).  These contaminants must be removed  
before we can derive a luminosity for the intracluster stars.  

Ultimately, the only definitive way to separate the two classes 
of objects is to perform follow-up spectroscopy: true IPN must 
also have the other emission lines found for planetary nebulae, 
the brightest being the [O~III] $\lambda$ 4959 line and the
Balmer lines.  So far, only a small fraction 
($\sim 30$) of the hundreds 
of IPN candidates have any follow-up spectroscopy at all, 
and of those, only a handful have enough signal-to-noise 
to see multiple emission lines in a single spectrum.   

However, we can statistically account for the contribution of the 
contaminants to IPN surveys by using control field techniques.  From 
a survey of 0.13 square degrees of blank regions far away from any galaxy 
cluster, and using identical procedures as our IPN surveys, 
we have found a total of nine IPN-like objects down to a 
limiting \m5007 magnitude of 27.0 (Ciardullo \etal 2002).  
When we compare this density to the density of our Virgo and 
Fornax surveys, we find a contamination rate of $\approx$ 20\% 
for Virgo and $\approx$ 50\% for Fornax.  The difference in 
contamination rate is primarily due to distance: Fornax is more distant 
than Virgo, and hence further down the contaminating source 
luminosity function.  

What are the contaminating objects?  In order for an object to 
masquerade as an IPN, it must have a very high observed 
equivalent width ($\geq 200$ \AA), because we require 
IPN candidates to be completely absent in our off-band exposures.  
From the equivalent widths, and spectroscopy (Kudritzki \etal 2000), 
the most likely source for the majority of the 
IPN contamination are Lyman-$\alpha$ galaxies at redshifts 
near 3.1, where the Lyman-$\alpha$ $\lambda$ 1215 line 
is Doppler-shifted into our [O~III] filter.  Other possibilities 
include type II quasars at high redshift (\eg Norman \etal 2002), 
and very rare high equivalent width [O~II] emitters 
at $z \approx 0.3$ (Stern \etal 2000).  These contaminants
are also now being found in normal PN galaxy surveys 
(\eg M{\' e}ndez \etal 2001).

\section{Luminosity Results}
After correcting for the three effects above, we can now derive the 
fraction of intracluster luminosity found from the IPN. 
After averaging the results from all of our fields, assuming a 
single distance for the IPN, a $\alpha_{2.5}$ value of 
$50 \times 10^{-9}$~PN-$L_{\odot}^{-1}$, and contamination rates 
of 20\% for Virgo and 50\% for Fornax, we find
that at least 20\% of all stars lie between the galaxies of Virgo and
Fornax.  The above assumptions are all conservative, in order 
to minimize the amount of intracluster starlight necessary 
to explain the data. 

The fractional results from the IRG observations are smaller than the
IPN results (10--20\%; Ferguson, Tanvir, \& von Hippel 1998; 
Durrell \etal 2002), but
are consistent within the errors.  Other intracluster light 
observations of more distant clusters (\eg Bernstein \etal 1995) 
find larger fractions for the amount of intracluster light, 
up to 50\%.  This could be due to systematic errors 
between the various methods, or due to differing galaxy cluster 
properties such as richness.  More data will be needed to resolve
this issue.

\section{Do IPN vary from ``normal'' PN?}

For most of this paper, we have focused on what the properties of 
PN imply about intracluster starlight.  What effects, 
if any, does the intracluster environment impose on the nebulae
themselves?

One possibility is that due to the large velocities ($v \sim 1000$ \kms) 
between the IPN and the intracluster medium, the expanding nebulae might 
fragment due to Rayleigh-Taylor instabilities (\eg Soker, 
Borkowksi, \& Sarazin 1991).  The nebulae would therefore 
be highly aspherical (\eg Villaver, Manchado, 
\& Garc{\' i}a-Segura 2000).  Unfortunately, it is unlikely 
that this effect could be observable.  There may be additional 
effects on the IPN, due to the hot intracluster medium, 
but detailed nebular modeling of the IPN is sorely needed. 

\section{The Future}

In the near future, there will be large samples of IPN candidates
in the nearby clusters of Virgo, Fornax, and Ursa Major, and
completed surveys in the much poorer M81 and Leo~I groups of 
galaxies.  Continued comparisons between IPN and IRG data sets
will be very useful.  As we have already shown, IPN and IRG observations
strongly complement each other: IPN surveys can map out 
the distribution and kinematics of the intracluster stars, while 
{\sl HST\/} IRG measurements at specific locations in clusters 
can provide strong constraints on the age and metallicity 
of the intracluster stars.

As the surveys continue, spectroscopic follow-up observations will
become critical to obtain velocities of the IPN, and to separate them
from the contaminating sources.  With deep observations
using 8-m class telescopes, we can obtain nebular abundances of the 
IPN, and compare them to PN of the Milky Way.

\vskip 0.125in

I would like to thank the organizing committee for inviting me
to give this presentation.  This work has been done in
collaboration with Robin Ciardullo, George Jacoby, Patrick
Durrell, Kara Krelove, and Steinn Sigurdsson.  
I would also like to thank Magda Arnaboldi, Ken Freeman, 
Rolf Kudritzki, \& Roberto M\'endez for their assistance 
and many useful discussions.  Travel support 
was given through the American Astronomical Society International 
Travel Grant.


\begin{references}
Arnaboldi, M., \etal 1996, \apj, 472, 145

Arnaboldi, M., \etal 2002, \aj, in press

Bernstein, G.M., \etal 1995, \aj, 110, 1507

Ciardullo, R. 1995, IAU Highlights of Astronomy, 10, 
ed. I. Appenzeller p.507

Ciardullo, R. 2002, \apj, in press

Ciardullo, R., Jacoby, G.H., \& Feldmeier, J.J. 1994, \baas, 185, 5216

Ciardullo, R., \etal 1998, \apj, 492, 62

Ciardullo, R., Jacoby, G.H., Ford, H.C., \& Neill, J.D. 
1989, \apj, 339, 53

Dressler, A. 1984, \araa, 22, 185

Durrell, P.R., \etal 2002, \apj, in press
 
Feldmeier, J.J. 2000, Ph.D. Thesis, Penn State University

Feldmeier, J.J., Ciardullo, R., \& Jacoby, G.H.\ 1998, \apj, 503, 109

Ferguson, H.C., Tanvir, N.R., \& von Hippel, T. 1998, Nature, 391, 461

Freeman, K.C., \etal 2000, in Galaxy Dynamics: From the Early 
Universe to the Present, 389.

Jacoby, G.H., Ciardullo, R., \& Ford, H.C. (JCF) 1990, \apj, 356, 332

Kudritzki, R.-P., \etal 2000, \apj, 536, 19

M\'{e}ndez, R.H., \etal 1997, \apj, 491, L23

M{\' e}ndez, R.H., \etal 2001, \apj, 563, 135

Moore, B., \etal 1996, Nature, 379, 613

Norman, C.A. \etal 2002, \apj, in press - available as astro-ph 0103198

Renzini, A. \& Buzzoni, A. 1986, in Spectral Evolution of Galaxies, 195

Soker, N., Borkowski, K.J., \& Sarazin, C.L.\ 1991, \aj, 102, 1381

Stern, D., Bunker, A., Spinrad, H., \& Dey, A.\ 2000, \apj, 537, 73

Theuns, T., \& Warren S.J. 1997, \mnras, 284, L11

Villaver, E., \etal 2000, Rev. Mex. Ast. Astrof. Conf. Ser., 9, 213

V{\'i}lchez-G{\'o}mez, R. 1999, in The Low Surface Brightness Universe, 349

Yasuda, N., Fukugita, M., \& Okamura, S. 1997, \apjs, 108, 417

Zwicky, F. 1951, \pasp, 63, 61

\end{references}
\end{document}